\documentclass[referee]{aa}
\usepackage{epsfig}
\usepackage{graphicx}

\begin{document}

\title{Forced oscillations of coronal loops driven by EIT waves}
\author{I. Ballai \inst{1}, M. Douglas\inst{1},  A. Marcu \inst{2}}
\institute{Solar Physics and Space Plasma Research Centre
(SP$^2$RC), Dept. of Applied Mathematics, University of Sheffield,
Hicks Building, Hounsfield Road, Sheffield, S3 7RH, England (UK)
\and Babes-Bolyai University, Dept. of Theoretical and
Computational Physics, 1 Kogalniceanu, 400084 Cluj-Napoca,
Romania}

   \offprints{I. Ballai, \email i.ballai@sheffield.ac.uk}

\authorrunning{Ballai et al.}
\titlerunning{Forced oscillations of coronal loops driven by EIT waves}
   \date{Received ; accepted }

\abstract {} {We study the generation of transversal oscillations
in coronal loops represented as a straight thin flux tube under
the effect of an external driver modelling the global coronal EIT
wave. We investigate how the generated oscillations depend on the
nature of the driver, and the type of interaction between the two
systems. } {We consider the oscillations of a magnetic straight
cylinder with fixed-ends under the influence of an external driver
modelling the force due to the global EIT wave. Given the
uncertainties related to the nature of EIT waves, we first
approximate the driver by an oscillatory force in time and later
by a shock with a finite width. }{Results show that for a harmonic
driver the dominant period in the generated oscillation belongs to
the driver. Depending on the period of driver, compared to the
natural periods of the loop, a mixture of standing modes harmonics
can be initiated. In the case of a non-harmonic driver (modelling
a shock wave), the generated oscillations in the loop are the
natural periods only. The amplitude of oscillations is determined
by the position of the driver along the tube. The full diagnosis
of generated oscillations is achieved using simple numerical
methods.}{}

 \maketitle

 \section{Introduction}

Latest high-resolution coronal observations have shown that
coronal structures are very dynamic entities with flows and waves
propagating along them. Waves and oscillations in coronal loops
have received increased attention in the last few years due to the
possibility to use the observed properties to diagnose not only
the magnetic field in these structures, but also the
sub-resolution space distribution of loops, plasma properties,
etc. (e.g. Roberts, Edwin and Benz 1984; Aschwanden et al. 1999;
and Nakariakov et al. 1999; Banerjee et al. 2007, Verth et al.
2007, Arregui et al. 2008, Verth and Er\'elyi 2008). One specific
type of oscillations observed in coronal loops is the fast kink
mode which disturbs the symmetry axis of the loop (for a full
description of these waves see, e.g. Edwin and Roberts 1983) and
it is nearly incompressible.

Many of kink waves and oscillations have their origin in the
interaction of coronal loops with various external sources and
drivers (see, e.g. Hindman and Jain 2008, Erd\'elyi and Hargreaves
2008). One of the possible explanations of oscillations in coronal
loops and/or prominence fibrils is that they have their origin in
the interaction of these loops with global coronal waves, e.g. EIT
waves. EIT waves (Thompson et al. 1999) are waves generated by
sudden energy releases (flares, coronal mass ejections, etc) and
they are able to propagate over very large distances in the solar
low corona.

Observational evidence for large-scale coronal impulses initiated
during the early stage of a flare and/or CME has been provided by
the EIT instrument onboard SOHO, TRACE/EUV, STEREO/EUVI. EIT waves
propagate in the quiet Sun with speeds of 250\,--\,400 km s$^{-1}$
at an almost constant altitude. At a later stage in their
propagation EIT waves can be considered as a freely propagating
wavefront which is observed to interact with coronal loops (see,
e.g. Wills-Davey and Thompson, 1999). Using TRACE/EUV 195 \AA\
observations, Ballai, Erd\'elyi and Pint\'er (2005) have shown
that EIT waves - seen in this wavelength - are waves with average
periods of the order of 400 seconds. Since at the height where the
EUV lines are formed, the magnetic field can be considered
vertical, EIT waves were interpreted as fast magnetohydrodynamic
(MHD) waves. This interpretation was confirmed using
multi-wavelength STEREO/EUVI observations by Long et al. (2008).
Recently, Attrill et al. (2007a,b) proposed that the diffuse EIT
coronal bright fronts are due to driven magnetic reconnections
between the skirt of the expanding CME magnetic field and the
favorably orientated quiet Sun magnetic field. According to this
latter model, the propagation process of the front consists of a
sequence of successive reconnection events.

Although a large consensus was reached on the trigger mechanism of
these global coronal waves and effects EIT waves can generate, the
nature of these large scale disturbances is still unknown, despite
the multitudes of models. The main reason of this uncertainty is
the lack of high temporal and spatial resolution as well as the
limited field of view (in the case of TRACE/EUV).

The present paper investigates the temporal and spatial variations
of transversal oscillations in a coronal loop under the influence
of an external driver representing the coronal global EIT waves.
{\bf Strictly speaking a coronal loop application would require
the consideration of an external magnetic field. However, we
consider this study as a starting point in a much more complex
analysis.} Due to the uncertainties in resolving the nature of EIT
waves, we will discuss separately the cases of a harmonic driver
and a driver of a finite width with a pulse-like temporal
distribution. It should be noted here that the study of the
present papers apply not only to EIT waves as external driver, but
it could be applied to any external source.

In the next section we introduce the working model and derive the
governing equation of transversal oscillations based on the
principle of force equilibrium in conjunction with the continuity
of mass and magnetic flux. The governing equation is solved for
the a {\bf driving force when its particular form is not
specified}. Section 3 is devoted to the study of the periodical
motion of the coronal loop under the influence of a few particular
drivers. Finally, in Section 4 we summarize our results and
discuss some possible key extensions which were neglected here but
they could be added to this model in subsequent studies.

 \section{Wave equations and solutions}

Let us suppose that a flux tube is situated in a magnetic free
 environment and it is under the effect of gravity. This model could approximate (in the first order) a magnetic loop
 in the solar corona. We consider
 the magnetic rope in the thin-flux-tube approximation inclined at an arbitrary angle with
 respect to the vertical. Let us suppose the
 directions ${\bf l}$ and ${\bf t}$ to be oriented along the loop
 and in a transversal direction. We assume that the EIT wave is
 acting with a force ${\bf F}$ (per unit volume) on the tube and the direction of
 this force is directed along the $x$-axis. For the sake of
 simplicity all dissipative effects are neglected. A schematic representation of
 the model is shown in Figure 1.
 \begin{figure}
   \centering
   \includegraphics[width=8.0cm]{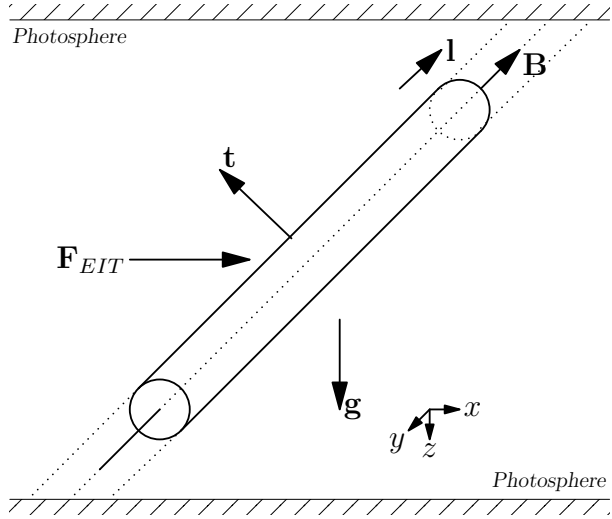}
      \caption{A schematic representation of the working model. The straight flux tube, arbitrarily
      inclined with respect to the
      vertical direction, is under the influence of different forces which will generate transversal oscillations in the
      coronal loop which has fixed ends in the dense photosphere.}
         \label{Figure 1}
   \end{figure}
Part of the discussion of this paper is using the model and
derivation developed by Spruit (1981) where the transversal waves
were studied in a magnetic flux tube in the convective
zone/photosphere.

 Any vector ${\bf a}$ can be decomposed with respect to the
 parallel and perpendicular direction of the tube such as
 \[
 {\bf a}_{\parallel}=({\bf \hat{l}}\cdot {\bf a}){\bf \hat{l}},
 \quad {\bf a}_{\perp}=({\bf \hat{l}}\times {\bf a})\times{\bf \hat{l}}
 \]
 The forces acting on the tube are the pressure force, the Lorentz
 force, the gravitational force and the force from the EIT wave. These
 forces are going to be decomposed along the two characteristic
 directions. We suppose that the homogeneous magnetic field is untwisted and
 it has a single component, along the tube, i.e. ${\bf B}=B{\bf
 \hat{l}}$.

 The parallel component of the motion is driven by the parallel
 components of acting forces. Since the Lorentz force has no field
 aligned component, it will appear only in the perpendicular
 direction. The parallel force equilibrium requires that
 \begin{equation}
 \rho_i\left(\frac{d {\bf v}}{d
 t}\right)_{\parallel}=-\partial_lp+\rho_i{\bf g}\cdot {\bf
 \hat{l}}+{\bf F}\cdot{\bf \hat{l}}
 \label{eq:2.1}
 \end{equation}
 where the operator $\partial_l$ is defined as $\partial_l={\bf
 \hat{l}}\cdot\nabla$ and $\rho_i$ is the density inside the loop.
 Along the perpendicular direction the forces acting on the tube
 will be
 \begin{equation}
 {\bf F}_{\perp}=-\left[{\bf \hat{l}}\times\nabla
 \left(p+B^2/\mu\right)\right]\times{\bf
 \hat{l}}+\left(\frac{({\bf B}\cdot\nabla){\bf
 B}}{\mu}\right)_{\perp}+\rho_i({\bf
 \hat{l}}\times{\bf g})\times{\bf
 \hat{l}}+({\bf
 \hat{l}}\times {\bf F})\times{\bf
 \hat{l}}
 \label{eq:2.2}
 \end{equation}
 According to Spruit (1981), the Lorentz force can be simply
 written as
 \begin{equation}
\left(\frac{({\bf B}\cdot\nabla){\bf
 B}}{\mu}\right)_{\perp}=\frac{B^2}{\mu}{\bf t}.
 \label{eq:2.3}
 \end{equation}
 We suppose that the tube is in equilibrium with its environment
 and the total pressure inside the tube is balanced by the
 external pressure, i.e.
 \begin{equation}
 \nabla(p+B^2/\mu)=\nabla p_e=\rho_e{\bf g},
 \label{eq:2.4}
 \end{equation}
 where $p_e$ and $\rho_e$ are the pressure and density outside the
 tube. With this in mind, the perpendicular component of the
 forces acting on the tube becomes
 \begin{equation}
{\bf F}_{\perp}=\frac{B^2}{\mu}{\bf t}+(\rho_i-\rho_e)({\bf
 \hat{l}}\times{\bf g})\times{\bf
 \hat{l}}+({\bf
 \hat{l}}\times {\bf F})\times{\bf
 \hat{l}}.
 \label{eq:2.5}
 \end{equation}
The perpendicular component of forces acts to move the plasma mass
(per unit volume) of $(\rho_i+\rho_e)$ in the tube and in the
exterior. Therefore the equilibrium of forces in the transversal
direction can be simply given as
\begin{equation}
(\rho_i+\rho_e)\left(\frac{d {\bf v}}{d
t}\right)_{\perp}=v_A^2\rho_i{\bf t}+(\rho_i-\rho_e)({\bf
 \hat{l}}\times{\bf g})\times{\bf
 \hat{l}}+({\bf
 \hat{l}}\times {\bf F})\times{\bf
 \hat{l}},
 \label{eq:2.6}
 \end{equation}
where we introduced the internal Alfv\'en speed
$v_A^2=B^2/(\mu\rho_i)$. It should be mentioned that in the
original derivation by Spruit (1981) the appearance of the term
containing $\rho_e$ on the left hand side of Eq. (\ref{eq:2.6})
was attributed to the apparent increase of tube's inertia.
Combining Eqs. (\ref{eq:2.1}) and (\ref{eq:2.6}), the total
equation of motion is given by
\begin{equation}
\frac{d {\bf v}}{d t}=-\frac{1}{\rho_i}\partial_lp{\bf
\hat{l}}+\left({\bf g}\cdot {\bf
 \hat{l}}\right){\bf
 \hat{l}}+\frac{1}{\rho_i}\left({\bf F}\cdot{\bf \hat{l}}\right){\bf
 \hat{l}}+\frac{\rho_i}{\rho_i+\rho_e}v_A^2{\bf t}+\frac{\rho_i-\rho_e}{\rho_i+\rho_e}({\bf
 \hat{l}}\times{\bf g})\times{\bf
 \hat{l}}+\frac{1}{\rho_i+\rho_e}({\bf
 \hat{l}}\times {\bf F})\times{\bf
 \hat{l}},
\label{eq:2.7}
\end{equation}
Let us write the Cartesian components of the unit vectors ${\bf
\hat{l}}=(l_x,l_y,l_z)$ and ${\bf \hat{t}}=(t_x,t_y,t_z)$, so the
cartesian components of Eq. ({\ref{eq:2.7}) can be written as
\[
\left(\frac{d {\bf v}}{d
t}\right)_x=-\frac{1}{\rho_i}\partial_lpl_x+gl_xl_z+\frac{\rho_i}{\rho_i+\rho_e}v_A^2t_x-\frac{\rho_i-\rho_e}{\rho_i+\rho_e}gl_xl_z
+\frac{F}{\rho_i+\rho_e}+
\frac{F\rho_e}{\rho_i(\rho_i+\rho_e)}l_x^2
\]
\[
\left(\frac{d {\bf v}}{d
t}\right)_y=-\frac{1}{\rho_i}\partial_lpl_y+gl_yl_z+\frac{\rho_i}{\rho_i+\rho_e}v_A^2t_y-\frac{\rho_i-\rho_e}{\rho_i+\rho_e}gl_yl_z+\frac{F\rho_e}
{\rho_i(\rho_i+\rho_e)}l_xl_y
\]
\begin{equation}
\left(\frac{d {\bf v}}{d
t}\right)_z=-\frac{1}{\rho_i}\partial_lpl_z+gl_z^2+\frac{\rho_i}{\rho_i+\rho_e}v_A^2t_z+\frac{\rho_i-\rho_e}
{\rho_i+\rho_e}g(1-l_z^2)+\frac{F\rho_e}{\rho_i(\rho_i+\rho_e)}l_xl_z
\label{eq:2.8}
\end{equation}
These equations must be supplemented by the continuity and
induction equations which can be combined into a single equation
of the form
\begin{equation}
\frac{d}{dt}\left(\frac{\rho}{B}\right)+\frac{\rho}{B}(\partial_l{\bf
v}_l+{\bf v}\cdot {\bf t})=0, \label{eq:2.9}
\end{equation}
where $v_l={\bf v}\cdot {\bf
 \hat{l}}$ and ${\bf t}=\partial_l {\bf
 \hat{l}}$. Now suppose that the flux tube is nearly
vertical and let us denote the horizontal displacement of the tube
by $\xi(z,t)$. In order to simplify the mathematics we suppose
that these displacements are small. According to Spruit (1981) the
components of the unit vector ${\bf \hat{l}}$ and ${\bf
 \hat{t}}$ can be written
as
\[
l_x=\frac{\partial \xi}{\partial z}, \quad l_z=1+{\cal O}(\xi^2),
\]
\begin{equation}
t_x=\frac{\partial^2\xi}{\partial z^2}+{\cal O}(\xi^2), \quad
t_z={\cal O}(\xi^2). \label{eq:2.10}
\end{equation}
In addition we assume that the tube is in the $xz$-plane, so we
choose $l_y=t_y=0$. Now the remaining two equations of the system
(\ref{eq:2.8}) reduce to
\begin{equation}
\left(\frac{d {\bf v}}{d
t}\right)_x=\left(-\frac{1}{\rho_i}\frac{\partial p}{\partial
z}+g-\frac{\rho_i-\rho_e}{\rho_i+\rho_e}g\right)\frac{\partial
\xi}{\partial
z}+\frac{\rho_iv_A^2}{\rho_i+\rho_e}\frac{\partial^2\xi}{\partial
z^2}+\frac{F}{\rho_i+\rho_e}+\frac{F\rho_e}{\rho_i(\rho_i+\rho_e)}\left(\frac{\partial
\xi}{\partial z}\right)^2, \label{eq:2.11}
\end{equation}
and
\begin{equation}
\left(\frac{d {\bf v}}{d
t}\right)_z=-\frac{1}{\rho_i}\frac{\partial p}{\partial
z}+g+\frac{F\rho_e}{\rho_i(\rho_i+\rho_e)}\frac{\partial
\xi}{\partial z}. \label{eq:2.12}
\end{equation}
In the above equations we restricted ourselves to linear motion
only. Next we assume that the vertical displacements are small and
of the same order as $\xi$. In order to obtain a closed equation,
in addition, we assume that the force, $F$ acting externally on
the tube is of the same order as $\xi$. Collecting terms of the
same order (with respect to $\xi$) in the two equations we obtain
that
\begin{equation}
\left(\frac{d {\bf v}}{d t}\right)_x=\frac{dv_x}{dt}+{\cal
O}(\xi^2)=\frac{\partial^2\xi}{\partial t^2}+{\cal O}(\xi^2),
\quad \frac{\partial p}{\partial z}=\rho_ig+{\cal
O}(\xi^2).\label{eq:2.14}
\end{equation}
After inserting these two relations back into Eq. (\ref{eq:2.11})
we obtain
\begin{equation}
\frac{\partial^2\xi}{\partial
t^2}=-\frac{\rho_i-\rho_e}{\rho_i+\rho_e}g\frac{\partial
\xi}{\partial
z}+\frac{\rho_iv_A^2}{\rho_i+\rho_e}\frac{\partial^2\xi}{\partial
z^2}+\frac{F}{\rho_i+\rho_e}. \label{eq:2.15}
\end{equation}
The above equation describes the propagation of transversal
oscillations of a vertical flux tube when the oscillations are
driven by an external force, $F$. Similar to Spruit (1981), the
first term on the right hand side is due to stratification and is
proportional to the buoyancy force, while the second term is due
to the restoring force due to the magnetic tension in the tube.
This equation (without the external force) has been originally
derived by Lamb (1932). Eq. (\ref{eq:2.15}) is similar to the
equation derived by Spruit (his equation 29) apart from the
driving term on the right hand side of our equation. The
propagation of kink modes described by a KG equation was studied
earlier by, e.g. Musielak and Ulmschneider (2003), Erd\'elyi and
Hargreaves (2008) and Hargreaves (2008).

In what follows we are going to solve Eq. (\ref{eq:2.15}) for a
coronal loop when the driving force is due to the incident EIT
wave. Before presenting the solutions we introduce a new function,
$Q$, defined as $\xi=Q\exp[\lambda z]$ and we choose the value of
the parameter $\lambda$ such that all first derivatives with
respect to $z$ vanish. After a straightforward calculation we
obtain that when
\[
\lambda=g\frac{\rho_i-\rho_e}{2\rho_iv_A^2}
\]
the governing equation (\ref{eq:2.15}) reduces to
\begin{equation}
\frac{\partial^2Q}{\partial
t^2}-\frac{\rho_iv_A^2}{\rho_i+\rho_e}\frac{\partial^2Q}{\partial
z^2}+g^2\frac{(\rho_i-\rho_e)^2}{4(\rho_i+\rho_e)\rho_iv_A^2}Q=\frac{Fe^{-\lambda
z}}{\rho_i+\rho_e}, \label{eq:2.16}
\end{equation}
which is a nonhomogeneous Klein-Gordon (KG) equation. A
nonhomogeneous KG equation has been also derived earlier by Rae
and Roberts (1982) and the inhomogeneous part described the effect
of the external medium. In their analysis the inhomogeneous part
was neglected by considering a situation where the temporal
variations of the parameters outside the loop are very slow
compared to changes inside the tube. For kink modes, Roberts
(2004) has obtained a similar equation.

The transversal waves described by equation (\ref{eq:2.16}) will
propagate with the speed given by the second term on the left hand
side
\begin{equation}
c_K=\sqrt{\frac{\rho_i}{\rho_i+\rho_e}}v_A \label{eq:2.17}
\end{equation}
which is the {\it kink speed}. This quantity has been previously
discussed within the context of wave propagation in magnetic flux
tubes by, e.g. Edwin and Roberts (1983). The coefficient of the
third term has dimension of $s^{-2}$ and its square root is given
by
\begin{equation}
\omega_c=\frac{g}{2v_A}\sqrt{\frac{(\rho_i-\rho_e)^2}{(\rho_i+\rho_e)\rho_i}},
\label{eq:2.18}
\end{equation}
and constitutes the cut-off frequency for kink modes propagating
in coronal loops. For typical coronal parameters ($v_A=900$ km
s$^{-1}$, $\rho_i/\rho_e=10$) the cut-off frequency of kink
oscillations is about 0.13 mHz. With these new notations, Eq.
(\ref{eq:2.16}) becomes
\begin{equation}
\frac{\partial^2 Q}{\partial t^2}-c_K^2\frac{\partial^2Q}{\partial
z^2}+\omega_c^2Q={\cal F}, \label{eq:2.19}
\end{equation}
where ${\cal F}=Fe^{-\lambda z}/(\rho_i+\rho_e)$.

Employing a normal mode analysis ($Q\sim e^{i(\omega t-kz)}$) for
the homogeneous part of Eq. (\ref{eq:2.19}), the dispersion
relation of these linear waves is given as
\begin{equation}
\omega=\pm\sqrt{k^2c_K^2+\omega_c^2}. \label{eq:2.20}
\end{equation}
Due to the particular $k$-dependence of the dispersion relation
waves are dispersive, i.e. waves with larger wavelength (shorter
$k$) propagating faster. The group speed of these waves is given
as
\[
\partial \omega/\partial
k=\pm\frac{kc_K^2}{\sqrt{k^2c_K^2+\omega_c^2}},
\]
so, waves with smaller wave number will have smaller group speed,
the maximum of the group speed (at $k\rightarrow \infty$) being
$c_K$.

Equation (\ref{eq:2.19}) has been studied in the context of pulse
propagation in the solar photosphere and chromosphere (see, e.g.
Roberts and Webb 1978, Rae and Roberts 1982, Kalkofen et al. 1994,
Sutmann et al. 1998, Hassan and Kalkofen 1999, Musielak and
Ulmschneider 2001, 2003; Hindman and Jain 2008, Erd\'elyi and
Hargreaves 2008). The solution of the KG equation represents the
propagation of a wave with the speed $c_K$ followed by a wake
oscillating with the frequency $\omega_c$.

An extension of the KG equation has been discussed by Ballai et
al. (2006) when the dissipation (kinematic viscosity in their
analysis) modified the KG equation into a Klein-Gordon-Burgers
equation where the dissipative term was given as a mixed (space
and time) derivative.

In what follows we present an analytical solution to Eq.
(\ref{eq:2.19}) in the most general form and particular solutions
will be deducted. Let us suppose that the boundary and initial
conditions {\bf used for solving} Eq. (\ref{eq:2.19}) are given by
\[
Q(0,t)=Q(L,t)=0,
\]
\begin{equation}
Q(z,0)=u_1(z), \quad \frac{\partial Q}{\partial t}(z,0)=u_2(z).
\label{eq:2.21}
\end{equation}
First we apply the Laplace transform to Eq. (\ref{eq:2.19}) and we
obtain
\begin{equation}
c_K^2\frac{\partial^2 \Psi}{\partial
z^2}-(s^2+\omega_c^2)\Psi=-\Phi-su_1(z)-u_2(z), \label{eq:2.22}
\end{equation}
where $\Psi(z,s)$ and $\Phi(z,s)$ are the Laplace transforms of
the functions $Q(z,t)$ and ${\cal F}(z,t)$ defined as
\[
\Psi(z,s)=\int_0^{\infty}Q(z,t)e^{-st}dt,\quad
\Phi(z,s)=\int_0^{\infty}{\cal F}(z,t)e^{-st}dt.
\]
Given the nature of the boundary conditions we further apply a
finite Fourier sine transform defined as
\[
F_s[f(x)]=\frac{1}{L}\int_0^Lf(x)\sin\alpha_nx\;dx,
\]
where we introduced
\[
 \alpha_n=\frac{n\pi}{L}.
\]
After applying the Fourier sine transform we obtain
\begin{equation}
(s^2+\omega_c^2+c_K^2\alpha_n^2){\overline \Psi}(n,s)={\overline
\Phi}(n,s)+s{\overline u}_1(n)+{\overline u}_2(n), \label{eq:2.23}
\end{equation}
where the functions with an {\it overline} represent the Fourier
transformed functions. From Eq. (\ref{eq:2.23}) we obtain that
\begin{equation}
{\overline \Psi}(n,s)=\frac{{\overline \Phi}(n,s)+s{\overline
u}_1(n)+{\overline u}_2(n)}{s^2+\omega_c^2+c_K^2\alpha_n^2}.
\label{eq:2.24}
\end{equation}
Now we apply an inverse Fourier transform which results in
\[
\Psi(z,s)=\frac{2}{L}\sum_{n=1}^{\infty}\frac{\sin\alpha_nz}{s^2+\omega_c^2+c_K^2\alpha_n^2}\left[\int_0^L
{\overline
\Phi}(\zeta,s)\sin\alpha_n\zeta\;d\zeta+s\int_0^L{\overline
u}_1(\zeta)\sin\alpha_n\zeta\;d\zeta+\right.
\]
\begin{equation}
\left.\int_0^L{\overline
u}_2(\zeta)\sin\alpha_n\zeta\;d\zeta\right]. \label{eq:2.25}
\end{equation}
In order to obtain $Q(z,t)$, we need to apply an inverse Laplace
transform to the function $\Psi(z,s)$. When calculating this
transform we will take into account the results of the convolution
theorem, i.e.
\[
{\cal L}^{-1}\left\{{\overline f}(s){\overline
g}(s)\right\}={\overline f}\ast{\overline
g}=\int_0^tf(\tau)g(t-\tau)d\tau,
\]
as well as the inverse Laplace transforms of the quantities
\[
{\cal
L}^{-1}\left(\frac{1}{s^2+\omega_c^2+c_K^2\alpha_n^2}\right)=\frac{1}{\sqrt{\omega_c^2+
c_K^2\alpha_n^2}}\sin\sqrt{\omega_c^2+c_K^2\alpha_n^2}t=\frac{1}{\omega_n}\sin\omega_nt,
\]
and
\[
{\cal
L}^{-1}\left(\frac{s}{s^2+\omega_c^2+c_K^2\alpha_n^2}\right)=\cos\sqrt{\omega_c^2+c_K^2\alpha_n^2}t=\cos\omega_nt,
\]
with $\omega_n= \sqrt{\omega_c^2+c_K^2\alpha_n^2}$ being the
natural frequency of the loop and the mode corresponding to $n=0$
being the cut-off frequency. It is interesting to note that the
ratio of periods corresponding to the fundamental mode ($n=1$) and
the first harmonic ($n=2$) is given by
\begin{equation}
\frac{T_1}{2T2}=\frac12\sqrt{\frac{\omega_c^2+4c_K^2\pi^2/L^2}{\omega_c^2+c_K^2\pi^2/L^2}}.
\label{eq:2.25.1}
\end{equation}
Due to the presence of the cut-off frequency this period ratio is
not 1 but is slightly smaller, however it can approach the
observed period ratio (Verwichte et al. 2004, McEwan et al. 2006)
if the cut-off is made unrealistically high. For typical coronal
and loop conditions, the natural periods of a loop of $L=200$ Mm
and $c_K=1000$ km s$^{-1}$ are $400$ s, $200$ s, and $133$ s,
respectively while for a $300$ Mm loop these periods will be in a
ratio $600/300/200$. If the length is fixed at $200$ Mm and let
the kink speed to be $1100$ km s$^{-1}$ the periods of the first
three modes will be $364/182/121$. It should be pointed out that
Eq. (\ref{eq:2.25.1}) is similar to the findings in McEwan et al.
(2006) though their equation was written for slow standing modes
(see their equation 24). However, observers have not reported
harmonics for slow waves whereas reports on higher harmonics for
kink modes are in abundance.

In the light of these relations, the inverse Laplace transform of,
e.g. the first term in Eq. (\ref{eq:2.25}) will be of the form
\begin{equation}
{\cal L}^{-1}\left(\frac{{\overline
\Phi}(\zeta,s)}{s^2+\omega_n^2}\right)=\frac{1}{\omega_n}\int_0^t\Phi(\zeta,\tau)\sin\left(\omega_n(t-\tau)\right)d\tau.
\label{eq:2.26}
\end{equation}
Applying a term-by-term inversion to the function $\Psi(z,s)$
given by Eq. (\ref{eq:2.25}) we obtain
\[
Q(z,t)=\frac{2}{L}\left\{\sum_{n=1}^{\infty}\frac{\sin\alpha_nz}{\omega_n}\int_0^L\sin\alpha_n\zeta\;d\zeta
\int_0^t\Phi(\zeta,\tau)\sin\left(\omega_n(t-\tau)\right)d\tau+\right.
\]
\begin{equation}
\left.\sum_{n=1}^{\infty}\sin\alpha_nz\left[\cos\omega_nt\int_0^Lu_1(\zeta)\sin\alpha_n\zeta\;d\zeta+\frac{\sin\omega_nt}
{\omega_n}\int_{0}^Lu_2(\zeta)\sin\alpha_n\zeta\;d\zeta\right]\right\}.
\label{eq:2.27}
\end{equation}
The solution of the nonhomogeneous Klein-Gordon equation given by
Eq. (\ref{eq:2.27}) can be simplified once the forms of the
functions $u_1$ and $u_2$ are known. In what follows we will
discuss a few particular cases and will investigate the
possibility of generating oscillations in coronal loops triggered
by an incident wave modelling the coronal global EIT wave.

The simplest particular case is when we have zero initial
conditions, i.e. $u_1(z)=u_2(z)=0$, and
\begin{equation}
{\cal F}(z,t)=f(t)\delta(z-z_0), \label{eq:2.28}
\end{equation}
where $\delta(z)$ is the Dirac-{\it delta} function. In this case
the solution of the inhomogeneous KG equation is given by
\begin{equation}
Q(z,t)=\frac{2}{L^2}\sum_{n=1}^{\infty}\frac{\sin\alpha_nz\sin\alpha_nz_0}{\omega_n}
\int_0^tf(\tau)\sin\left(\omega_n(t-\tau)\right)d\tau,
\label{eq:2.29}
\end{equation}
where we used the property that
\[
\delta(z-z_0)\sin\alpha_nz=\left\{%
\begin{array}{ll}
    \sin\alpha_nz_0, & \hbox{if $0\leq z_0\leq L$} \\
    0, & \hbox{otherwise}. \\
\end{array}%
\right.
\]
When deriving Eq. (\ref{eq:2.29}) we took into account that the
${\delta}$-function has a dimension of $L^{-1}$ and a variable
change of the form ${\tilde z}=z/L$ and ${\tilde z_0}=z_0/L$ is
needed. The extra $L$ in the denominator of Eq. (\ref{eq:2.29})
arises after we apply the property that $\delta[L({\tilde
z}-{\tilde z_0})]=L^{-1}\delta({\tilde z}-{\tilde z_0})$. If we
further assume that $f(\tau)=\delta(\tau)$ (i.e. the source
consists of an impulse acting at $z=z_0$) the solutions describing
the oscillations in a fixed-ends loop is given by
\begin{equation}
Q(z,t)=2c_KL\sum_{n=1}^{\infty}\frac{\sin\alpha_nz\sin\alpha_nz_0}{\omega_n}
\sin\omega_nt=G(z,t/z_0) \label{eq:2.30}
\end{equation}
which constitute the Green function for the coronal loop modelled
as a straight structure fixed at $z=0,L$. Once the Green function
is known, the solution of the inhomogeneous KG equation for an
arbitrary external action ${\cal F}(z,t)$ can be written as
\begin{equation}
Q(z,t)=\int_0^Ldz^\prime\int_0^tG\left(z,t-\tau/z^{\prime}\right){\cal
F}(z^\prime,\tau)d\tau. \label{eq:2.31}
\end{equation}
The present analysis does not include any information about the
radius of the tube geometrical or the internal structure of the
tube and external magnetic field, factors which could be
important. However, it is easy to estimate the magnitude of the
external force required to induce oscillations in the tube. The
magnetic tension force in the tube with constant circular
cross-section is $(B^2/\mu)\pi R^2$ where $R$ is the constant
radius of the tube. The external force must be {\it at least} as
large as the tension of the tube. Writing a simple force
equilibrium equation in transversal direction to the axis fo the
tube we obtain that the external force acting on the tube in a
point $z_0$ along the tube has to be larger than
\[
\frac{B^2}{\mu}\pi
R^2\left[\frac{1}{(1+\lambda_e^2z_0^2)^{1/2}}+\frac{1}{(1+\lambda_e^2(L-z_0)^2)^{1/2}}\right],
\]
where $1/\lambda_e$ is the maximum displacement of the tube and is
given by (see, e.g. Edwin and Roberts 1983)
\[
\lambda_e=k\sqrt{\frac{c_K^2-c_{Se}^2}{c_{Se}^2}},
\]
with $c_{Se}$ being the sound speed in the magnetic free region
outside the coronal loop and $k=\pi n/L$ is the longitudinal
wavenumber. For a loop length of $200$ Mm and $c_K=1000$ km
s$^{-1}$, $c_{Se}=200$ km s$^{-1}$ we obtain a maximum
displacement of the fundamental mode of $12.9$ Mm.

In the following section we are going to discuss a few particular
cases referring to the nature of the driver and find the equation
giving the transversal displacement of the loop as given by Eq.
(\ref{eq:2.27}).

\section{Drivers of particular form}

The discussion of these separate particular cases is needed as the
true nature of EIT waves is not known. As specified before, the
force on the right hand side of Eq. (\ref{eq:2.15}) is the force
which acts on the coronal loop and represents the effect of the
incident EIT wave on the coronal loop. Obviously it is difficult
to estimate the value (or the direction) of this force, however,
some estimations can be made (see also the end of the previous
section). If we suppose that the entire energy of the EIT wave
($E_{EIT}$) is converted into inducing oscillations of the loop,
the energy of the EIT wave will work toward displacing the loop.
Therefore we can write that
\[
E_{EIT}=\frac{F}{\lambda_e}.
\]
Obviously the energy of EIT waves is quantity which cannot be
directly measured however, previous indirect estimations (Ballai
2007) show that these energies are in the range of
$10^{16}-10^{19}$ J.

\subsection{Harmonic drivers}

Let us suppose that the EIT wave is a wave and its action of the
coronal loop is described by a force of the form
\begin{equation}
{\cal F}=E_{EIT}\lambda_e\frac{\delta(z-z_0)e^{-\lambda
z}e^{i\omega_{EIT}t}}{\rho_i+\rho_e}, \label{eq:2.32}
\end{equation}
where $\omega_{EIT}$ is the frequency of EIT waves. This form of
the externally acting force is inserted back into Eq.
(\ref{eq:2.27}), yielding
\[
Q(z,t)=-\frac{4E_{EIT}\lambda_e}{L^3(\rho_i+\rho_e)}\sum_{n=1}^{\infty}\frac{e^{-\lambda
z_0}\sin\alpha_nz\sin\alpha_nz_0}
{(\omega_{EIT}^2-\omega_n^2)}\left\{
\left[\sin\left(\frac{\omega_n+\omega_{EIT}}{2}\right)t\times\right.\right.
\]
\begin{equation}
\left.\left.\sin\left(\frac{\omega_n-\omega_{EIT}}{2}\right)t\right]+i(\omega_{EIT}\sin\omega_nt-
\omega_n\sin\omega_{EIT}t)\right\}. \label{eq:2.33}
\end{equation}
The presence of the $\sin\alpha_nz_0$ in the expression $Q(z,t)$
simply means that the amplitude of generated oscillations will
depend on the height (along the loop) where the EIT wave interacts
with the loop. The maximum amplitude of oscillations (in the case
of modes with odd $n$) will be reached when the EIT wave hits the
top of the loop, i.e. the interaction occurs at $z_0=L/2$. In this
case there will be no modes generated corresponding to an even $n$
(for instance for an interaction of this type we will not have
first harmonics present in the loop). This statement supports the
conclusions drawn by Ballai (2007) where a list with possible
factors which can influence the characteristics of loop
oscillations was given. The singularity in the denominator of Eq.
(\ref{eq:2.33}) is just apparent, its effect is balanced by the
numerator. If we concentrate only on the real part of Eq.
(\ref{eq:2.33}), it is also obvious that the resulting signal will
not have a well defined standing mode pattern, instead the
oscillations will be a superposition of different oscillations.
The real part of Eq. (\ref{eq:2.33}) is numerically represented in
Figure 2 (in all subsequent figures we will always use the real
part). The period of the driver EIT wave is left to vary between
50 and 800 seconds.

 \begin{figure}
   \centering
   \includegraphics[width=8.0cm]{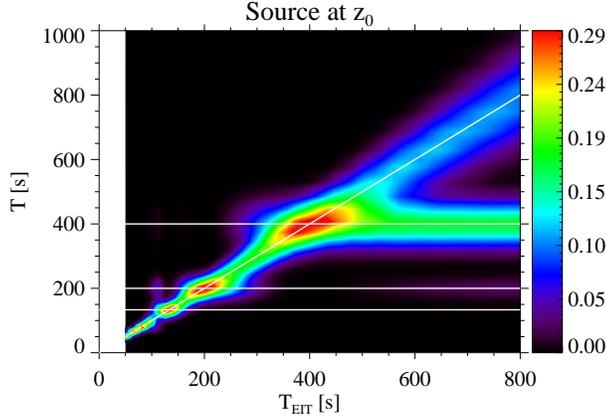}
      \caption{The periods of oscillations generated by an EIT wave acting at $z=z_0$. The period of the
      driver is changed in the interval $50$ to $800$ s. The horizontal lines represent the natural
      periods of the loop, i.e. $T_n=2\pi/\omega_n$. The inclined line corresponds to the periods of the EIT wave,
      the driver of the oscillations in the coronal loop.}
         \label{Figure 2}
   \end{figure}
The pattern of oscillations which can be generated in the coronal
loop depends on the characteristics of the driver. For each period
of the driver (in between 50 and 800 s) a wavelet analysis has
been carried out for the temporal part of Eq. (\ref{eq:2.33}). The
power of the signal has been summed up inside the cone of
influence set at a confidence level of 95\% and the results are
shown in Figure 2 which is color-coded, the red color corresponds
to the highest power, while the black color represents the lowest
power. Depending on the period of the EIT wave, various
oscillation modes can be excited. For the example shown in Figure
2, the loop has a length of $200$ Mm and a kink speed of $1000$ km
s$^{-1}$. The natural periods corresponding to these values are in
a ratio of $400/200/133$, values represented by the horizontal
lines.

Let us consider a driver which has a period larger than $600$ s.
In this case, the modes which can be excited will be the
fundamental mode (corresponding to $400$ s) and the first harmonic
but with a very low power. The oscillation pattern of the driver
is still present but much weaker than the oscillation of the
fundamental mode. As the period of the driver becomes smaller,
other harmonics can be excited. For a period between $200$ and
$400$ s the pattern of the driver is preserved (see the inclined
bright direction) but a considerable amount of the fundamental
mode and first harmonic can be generated. Higher harmonics are
also present but their power is very small. For a period of less
than $200$ s the dominant oscillations will be the first and
second harmonics, while the fundamental mode is extremely weak.
The red regions correspond to the cases when the period of the
driver matches (or is very close) to one of the natural periods of
the loop. In that case there is a resonance between the driver and
the coronal loop.

In reality, however, if the EIT wave is an oscillating front
colliding with the coronal loop, then the interaction occurs not
only in one point, but in two, symmetrically situated from the
ends of the loop. Let us suppose now that the driver is a front
which interacts with the loop {\it at the same time} in two
points, at $z_0$ and at $L-z_0$. In this case, the driver will
have the form
\[
{\cal
F}=E_{EIT}\lambda_e\frac{[\delta(z-z_0)+\delta(z-L+z_0)]e^{-\lambda
z}}{\rho_i+\rho_e}e^{i\omega_{EIT}t}.
\]
For this expression the resulting oscillations will be described
by a similar function as given by Eq. (\ref{eq:2.33}) but now the
spatial dependence will contain in the numerator the expression
\[
e^{\lambda z_0}\sin
\alpha_nz_0+e^{-\lambda(L-z_0)}\sin\alpha_n(L-z_0).
\]
This situation can be achieved if the front of the EIT wave is
perfectly perpendicular to the coronal loop. For this particular
driver the corresponding period-diagram is shown in Figure 3.
\begin{figure}
   \centering
   \includegraphics[width=8.0cm]{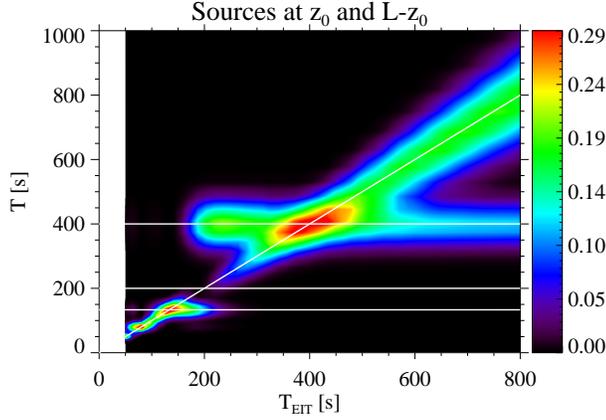}
      \caption{The same as in Figure 2 but now the oscillations are generated by an
      EIT wave interacting with the coronal loop at $z=z_0$ and $z=L-z_0$. }
         \label{Figure 3}
   \end{figure}
According to expectations, in this case no modes corresponding to
an even $n$ will be excited. In contrast to the first case, for
driver's period larger than the period corresponding to the first
natural period ($400$ s) the EIT wave will excite modes which will
carry predominantly the characteristic of the driver and in a
smaller quantity the properties of the fundamental mode. No first
harmonic can be generated, instead for a narrow range of the
driver's period, (a small interval around $200$ s) the
oscillations will comprise addition from the fundamental mode and
the second harmonic.

In reality it is more likely that the front of the incident EIT
wave is not completely perpendicular to the axis of the loop, now
the two interaction points between the loop and EIT wave will
separated by a delay time, i.e. the time required for the front to
reach the other half of the loop. The delay time can be easily
calculated (see for details Ballai 2007) and depends on the length
of the loop, the speed of propagation of the EIT wave and the
attack angle, i.e. the angle the front of the EIT wave makes with
the vertical plane of the coronal loop. In this case, the acting
force will have a spatial and temporal dependence of the form
\[
\delta(z-z_0)e^{i\omega_{EIT}t}+\delta(z-L+z_0)e^{i\omega_{EIT}(t-T_d)}H(t-T_d),
\]
where $T_d$ is the delay time and $H(t)$ is the Heaviside step
function. After inserting this form back into Eq. (\ref{eq:2.27})
we obtain that the modified transversal displacement of the
coronal loop is of the form
\[
Q(z,t)=-\frac{4E_{EIT}\lambda_e}{L^3(\rho_i+\rho_e)}\sum_{n=1}^{\infty}\frac{\sin\alpha_nz}
{(\omega_{EIT}^2-\omega_n^2)}\left\{e^{-\lambda
z_0}\sin\alpha_nz_0\left[\cos\omega_nt-\cos\omega_{EIT}t\right]-
\right.
\]
\begin{equation}
-\left.e^{-\lambda(L-z_0)}\sin\alpha_n(L-z_0)\left[\cos\omega_n(t-T_d)-\cos\omega_{EIT}(t-T_d)\right]\right\}.
\label{eq:2.34}
\end{equation}
As a particular case we have chosen the situation when the delay
time corresponds to an integer number of EIT wave's period. In the
case of two external forces acting in phase upon the coronal loop
the possible modes appearing in the coronal loop are shown in
Figure 4.
\begin{figure}
   \centering
   \includegraphics[width=8.0cm]{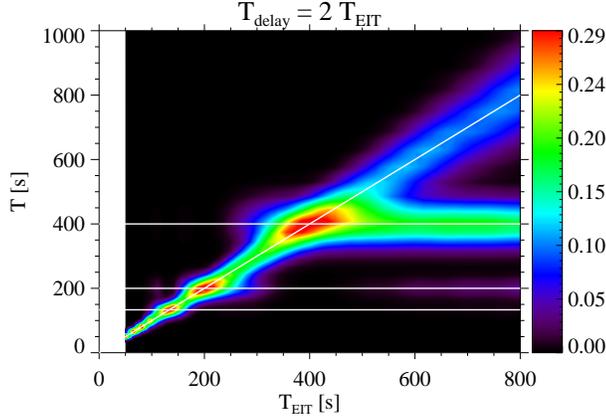}
      \caption{The same as in Figure 3 but now the oscillations are generated by an EIT
      wave acting at $z=z_0$ and $z=L-z_0$ and the second interaction is delayed by a time corresponding to the
      double of EIT waves' period. }
         \label{Figure 4}
   \end{figure}
It is obvious that the period of generated oscillations will
contain the period of the EIT wave as the strongest component. For
periods larger than the natural period of the fundamental mode the
generated oscillation will be dominated by the period of the
fundamental mode with a weaker signal resembling the
characteristics of the driver and a very weak period corresponding
to the first harmonic. For periods of the driver situated between
the natural periods of the loop, the possibility of mode
generation goes parallel with the case explained in the case of a
single driver. It can be easily shown that the distribution of
possible periods in the coronal loop is similar even when the
delay time is not an integer number of EIT waves' period.

\subsection{Non-harmonic driver}

EIT waves have also been explained in terms of a non-wave feature
(i.e. not having a harmonic behaviour). In this context we could
list the models proposed by, e.g. Delan\'ee (2000) and Attrill et
al. (2007) where EIT waves were associated with deformations and
evolutions of magnetic fields resulted after the release of the
CME. In order to include this possible explanation of EIT waves,
let us suppose an external force acting on the coronal loop of the
form
\begin{equation}
{\cal
F}=E_{EIT}\lambda_e\frac{e^{-\lambda_z}\left[H(z-z_0)-H(z-z_0^\prime)\right]}{\rho_i+\rho_e}\delta(t),
\label{eq:2.35}
\end{equation}
which means that the external driving force is represented by a
finite width ($|z_0-z_0^\prime|$) front which has no temporal
component other than a Dirac-delta function. If the form of
external force given by Eq. (\ref{eq:2.35}) is inserted back into
Eq. (\ref{eq:2.27}), we obtain that the temporal part of the
transversal displacement of the magnetic tube modelling a coronal
loop is given by
\[
\int_0^t\delta(t^\prime)\sin\omega_n(t-t^\prime)dt^\prime\sim\sin\omega_nt,
\]
which means that this form of the EIT wave (a single finite width
front) will produce oscillations in the coronal loop at the
natural frequency of the loop only.

It is possible that the two forms of the external driver (the
harmonic and non-harmonic) treated here coexist in the sense that
they are the manifestation of the same phenomenon but at different
distance from the source, therefore a more careful analysis will
be needed in the future.

 \section{Conclusions}

 The generation and propagation of oscillations in coronal loops
 modelled by a straight magnetic cylinder with fixed ends is
 studied when the coronal loop is under the effect of an EIT wave, as
 a driver. We found that if all forces acting on the flux tube are taken into
 account, the governing equation describing the propagation of
 standing transversal (kink) waves is described by an inhomogeneous
 Klein-Gordon-type equation and the inhomogeneous part of the
 equation is represented by the force (over a unit volume)
 by EIT waves. The evolutionary equation contains information
 about the propagation speed of waves (here the kink speed) and
 the cut-off frequency of kink modes. The cut-off value is
 determined by the densities inside/outside the loop and the
 Alfv\'en speed (i.e. magnetic field).

 Using the combined Laplace and Fourier sine transform techniques, the governing
 equation is solved, such that the solution takes into
 account general initial and boundary conditions.

 Particular solutions have been found in the case of an EIT wave
 considered first as a wave with a frequency $\omega_{EIT}$, and later
 as a shock wave with finite front thickness (a non-harmonic
 driver). The results show that in case of a non-harmonic driver
 the periods of generated modes always belong to the natural periods of the loop.
 On the other hand, in the case of a periodic driver - for
 an arbitrary period of the driver - there will be a mixture of standing modes which
 could explain on the observed period ratio. The
 analysis carried out here for different type of drivers show
 that the generated oscillations will carry predominantly
 information about the driver rather than the loop itself.

 The oscillations described in this paper were all modelled in the
 framework of ideal MHD. In reality coronal loop oscillations are
 observed to decay relatively rapidly and several mechanisms have
 been proposed to explain this damping (Ruderman and Roberts 2005,
 Terradas et al. 2005, 2007, Selwa et al. 2007).
 The inclusion of a
 dissipative (or energy lost) mechanism in the present model will
 be addressed in a future study. It could be possible that in the
 case of a loop oscillating as a whole in the kink mode, the friction with the
 environment could be also an important factor whose inclusion in
 the model could result in a possible explanation of the damping
 of these oscillations.

 The present study can be further extended to the case when the
 external EIT wave acts not only on a single magnetic loop but on
 a system of adjacent loops. In this case the primary displacement of the first coronal loop
 (generated by the incoming EIT wave) will be the driver for the
 oscillations in the second loop (and so on), leading to coupled loop oscillations.
 {\bf In order to describe a realistic loop, the present
 model is going to be expanded to consider the effect of an
 external magnetic field. It is expected that the presence of this
 field will generate an additional force which will tend to suppress
 the oscillation of the tube.}

 \begin{acknowledgements} I.B. acknowledges the financial support by
NFS Hungary (OTKA, K67746). IB and AM were supported by The
National University Research Council Romania
(CNCSIS-PN-II/531/2007). M.D. acknowledges the support from STFC.
\end{acknowledgements}


\begin{thebibliography}{}
\bibitem[\protect\citeauthoryear{Arregui et al.}{2008}]{are08}
Arregui, I., Terradas, J., Oliver, R. \& Ballester, J.L. 2008,
\apj, 674, 1179
\bibitem[\protect\citeauthoryear{Aschwanden et al.}{1999}]{asc99}
Aschwanden, M.J., Fletcher, L., Schrijver, C.J., Alexander, D.
1999, \apj, 520, 880
\bibitem[\protect\citeauthoryear{Attrill et al.}{2007}]{att07a} Attrill, G.D.R., Harra, L.K., van Driel-Gesztelyi, L.,
D\'emoulin, P. \& W\"usler, J.-P. 2007a, Astron. Nachrt., 328, 760
\bibitem[\protect\citeauthoryear{Attrill et al.}{2007}]{att07b} Attrill, G.D.R., Harra, L.K., van Driel-Gesztelyi,
L.\& D\'emoulin, P. 2007b, \apj, 656, L101
\bibitem[\protect\citeauthoryear{Ballai}{2007}]{bal07}
Ballai, I. 2007, \solphys, 247, 177
\bibitem[\protect\citeauthoryear{Ballai et al.}{2006}]{bal06} Ballai, I.,
Erd\'elyi, R., Hargreaves, J. 2006, Phys. Plasmas, 13, 042108
\bibitem[\protect\citeauthoryear{Ballai, Erd\'elyi and Pint\'er}{2005}]{bal05}
Ballai, I., Erd\'elyi, R., Pint\'er, B. 2005, \apj, 633, L145
\bibitem[\protect\citeauthoryear{Banerjee et al.} {2007}]{ban07} Banerjee, D., Erd\'elyi, R., Ramon, O. \& O'Shea, E. 2007,
\solphys, 246, 3
\bibitem[\protect\citeauthoryear{Delan\'ee} {2000}]{del00} Delan\'ee, C. 2000, \apj, 545,
512
\bibitem[\protect\citeauthoryear{Edwin and Roberts}{1983}]{edw83}
Edwin, P. \& Roberts, B. 1983, \solphys, 88, 179
\bibitem[\protect\citeauthoryear{Erd\'elyi and Hargreaves}{2008}]{erd08}
Erd\'elyi, R. \& Hargreaves, J. 2008, \aap, (in press)
\bibitem[\protect\citeauthoryear{Hargreaves}{2008}]{har08}Hargreaves,J. 2008:
{\it PhD Thesis}, University of Sheffield, pp. 84
\bibitem[\protect\citeauthoryear{Hassan and Kalkofen}{1999}]{has99} Hasan, S.S.  \& Kalkofen, W. 1999, \apj, 519, 899
\bibitem[\protect\citeauthoryear{Hindman and Jain}{2008}]{hin08} Hindman, B.W. \& Jain, R. 2008, \apj, 677, 899
\bibitem[\protect\citeauthoryear{Kalkofen et al.}{1994}]{kal94} W. Kalkofen, W., Rossi, P.,
Bodo, G. \& Massaglia, S. 1994, \aap, 284, 976
\bibitem[\protect\citeauthoryear{Lamb}{1932}]{lamb32}Lamb,H. 1932,
{\it Hydrodynamics}, Cambridge University Press, pp. 138
\bibitem[\protect\citeauthoryear{Long et al.}{2008}]{lon08}
Long, D.M., Gallagher, P.T., McAteer, R.T.J. \& Bloomfield, D.S.
2008, \apj, 680, 81
\bibitem[\protect\citeauthoryear{Rae and Roberts}{1982}]{rae82} Rae, I.C. \& Roberts, B. 1982, \apj, 256, 761
\bibitem[\protect\citeauthoryear{Roberts, Edwin and Benz}{1984}]{rob84} Roberts, B., Edwin, P.M., Benz, A.O. 1984,
\apj, 279, 857
\bibitem[\protect\citeauthoryear{Roberts and Webb}{1978}]{rob78} Roberts, B. \& Webb, A.R. 1978, \solphys, 56, 5
\bibitem[\protect\citeauthoryear{Roberts}{2004}]{rob04} Roberts,
B., in {\it Waves, oscillations and small-scale transient events
in the solar atmosphere: a joint view from SOHO and TRACE}, 2004,
ESA SP-547, 1
\bibitem[\protect\citeauthoryear{Ruderman and Roberts}{2002}]{rud02} Ruderman, M.S. \& Roberts, B. 2002, \apj, 577, 475

\bibitem[\protect\citeauthoryear{McEwan et al.}{2006}]{mce06}
McEwan, M.P., Donnelly, G.R., Diaz, A.J. \& Roberts, B. 2006,
\aap, 460, 893
\bibitem[\protect\citeauthoryear{Musielak and Ulmschneider}{2001}]{mus01} Musielak, Z.E. \& Ulmschneider, P. 2001, \aap, 370, 541
\bibitem[\protect\citeauthoryear{Musielak and Ulmschneider}{2003}]{mus03} Musielak, Z.E.
\& Ulmschneider, P. 2003, \aap, 400, 1057
\bibitem[\protect\citeauthoryear{Nakariakov et al.}{1999}]{nak99}
Nakariakov, V.M., Ofman, L., DeLuca, E.E., Roberts, B. \& Davila,
J.M. 1999, {\em Science}, 285, 862
\bibitem[\protect\citeauthoryear{Selwa et al.}{2007}]{sel07}
Selwa, M., Murawski, K., Solanki, S.K. \& Wang, T.J. 2007, \aap,
462, 1127
\bibitem[\protect\citeauthoryear{Spruit}{1981}]{spru81} Spruit,
H.C 1981, \aap, 98, 155
\bibitem[\protect\citeauthoryear{Sutmann et al.}{1998}]{sut98} Sutmann,
G., Musielak, Z.E. \& Ulmschneider, P. 1998, \aap, 340, 556
\bibitem[\protect\citeauthoryear{Terradas et al..}{2007}]{ter07}
Terradas, J., Andries, J. \& Goossens, M. 2007, \solphys, 246, 231
\bibitem[\protect\citeauthoryear{Terradas et al.}{2005}]{ter05}
Terradas, J., Oliver, R. \& Ballester, J.L. 2005, \aap, 441, 371
\bibitem[\protect\citeauthoryear{Thompson et al.}{1999}]{tho99} Thompson, B.J., Gurman, J.B., Neupert, W.M. et
al. 1999, \apj, 517, 151
\bibitem[\protect\citeauthoryear{Verth et al.}{2007}]{ver07}
Verth, G., van Doorselaere, T., Erd\'elyi, R. \& Goossens, M.
2007, \aap, 457, 341
\bibitem[\protect\citeauthoryear{Verth et al.}{2008}]{ver08}
Verth, G. \& Erd\'elyi, R. 2008, \aap, (in press)
\bibitem[\protect\citeauthoryear{Verwichte et al.}{2004}]{ver04}
Verwichte, E., Nakariakov, V.M., Ofman, L. \& Deluca, E.E. 2004,
\solphys, 233, 77
\bibitem[\protect\citeauthoryear{Wills-Davey and
Thompson}{1999}]{wil99} Wills-Davey, M.J. \& Thompson, B.J. 1999,
\solphys, 190, 467



\end{thebibliography}
 \end{document}